\documentclass[conference]{IEEEtran}
\advance\topmargin4pt
\advance\textheight-8pt

\usepackage[nospace]{cite}

\usepackage{mathtools, amssymb}
    \def\A{\mathsf A}

    \def\EE{\mathbb E}
    \def\PP{\mathbb P}
    \def\CCC{\mathcal C}
    \def\III{\mathcal I}
    \def\OOO{\mathcal O}

    \def\BEC{\operatorname{BEC}}

\usepackage{tikz, booktabs}
    \pgfmathdeclarerandomlist{ACGT}{{A}{C}{G}{T}}
    \tikzset{
        every picture/.style = {line cap=round,line join=round},
        A/.style={blue, fill=.!10},
        C/.style={teal, fill=.!10},
        G/.style={blue!30!teal, fill=.!10},
        T/.style={blue!50!teal, fill=.!10},
        base/.style={
            transform shape, above right,
            circle, draw, fill, inner sep=1pt,
        }
    }

\begin{document}

                                   \title
    {Salami Slicing Trellis for \\ Synchronization Errors in DNA Coding}
                                      
                                   \author
{Tsung-Han Wu, Joseph Swernofsky, Hsin-Po Wang (National Taiwan University)}
                                      
                                 \maketitle

\begin{abstract} \boldmath
    On top of substitution errors, DNA storage channels suffer from
    both insertions and deletions at the same time.
    It is therefore important to develop error-correcting codes with
    efficient encoders and decoders that can combat all three types of noise.
    This paper introduces the salami-slicing trellis,
    a decision-feedback trellis that computes bitwise posterior probabilities
    along each strand and is coupled with polar codes across strands.
    The decoder alternates between advancing the trellises by one position
    and polar-decoding the resulting cross-strand slice,
    feeding the decoded bits back to the trellises for the next position.
    Simulations suggest that the resulting coding scheme approaches
    the conjectured capacity of the substitution-insertion-deletion channel.
\end{abstract}

\section{Introduction}

    DNA has emerged as a promising medium for long-term data storage.
    With a storage density of 200 petabytes per gram---seven orders
    of magnitude higher than consumer-grade hard drives---DNA offers
    unprecedented potential for archival applications \cite{CGK12,GBC13}.
    Moreover, DNA can preserve information for centuries without external
    power~\cite{WPK23}, making it ideal for backup systems.
    These advantages have motivated startups such as Atlas, Catalog,
    DNA Script, Iridia, and DNAli to invest in DNA storage technology.

    Despite these advantages, DNA storage systems face unique challenges
    that distinguish them from traditional electromagnetic storage
    media \cite{SKS24,MiP24}.
    Current DNA sequencing technologies, such as nanopore
    sequencing~\cite{YHZ22}, introduce substitution, insertion,
    and deletion errors at rates of approximately 0.1\%--1\% \cite{SKK22}.
    Of these error types, substitution errors are well studied
    in classical coding theory,
    but insertions and deletions remain comparatively less explored.
    The simultaneous presence of all three error types creates
    a ``triply-noisy'' channel that requires nontrivial coding techniques.

\begin{figure}
    \centering
    \begin{tikzpicture}
        \foreach \x in {1, ..., 8} {
            \tikzset{xshift = \x * 1cm}
            \foreach \y in {1, ..., 4} {
                \pgfmathrandomitem\A{ACGT}
                \tikzset{yshift = \y * 0.5cm}
                \draw [\A] (0, 0.4) rectangle (0.1, 0)
                    node [base] {\A} (0.5, 0.4);
            }
            \draw [rounded corners=3pt] (-0.1, 0.4) rectangle (0.6, 2.5)
                (0.25, 0.4) node [below, align=center, scale=0.7]
                {inner \\ IDS \\ code} ;
        }
        \draw [rounded corners=12pt](0.6, -0.5) rectangle (9, 2.6);
        \draw (4.8, -0.5) node [below] {outer code (usually Reed--Solomon)};
    \end{tikzpicture}
    \caption{
        The concatenated code design:
        Each strand is protected by an inner code
        that can handle insertions, deletions, and/or substitutions.
        All strands together are protected by an outer code
        that is a Reed--Solomon code.
    }                                                      \label{fig:concat}
\end{figure}
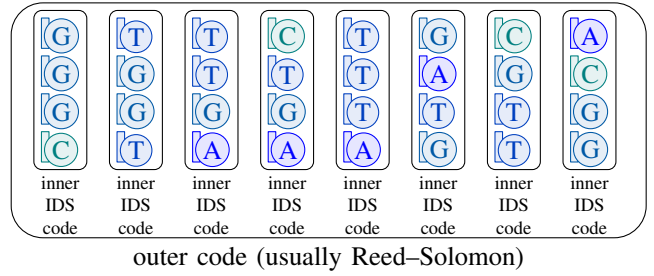

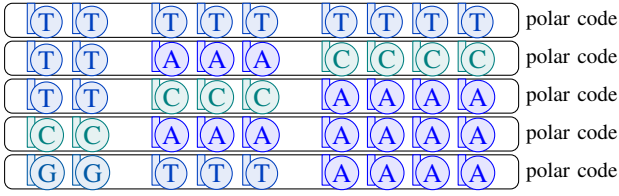
\begin{figure}
    \centering
    \begin{tikzpicture}
        \footnotesize
        \foreach \y in {1, ..., 5} {
            \tikzset{yshift = \y * 0.5cm}
            \draw [rounded corners=3pt] (1.8, -0.05) rectangle (8.6, 0.4)
                node [below right] {polar code};
        }
        \normalsize
        \foreach \x in {2, ..., 4} {
            \tikzset{xshift = \x * (\x+0.5) * 0.3cm}
            \foreach \xx in {1, ..., \x} {
                \pgfmathsetseed{20260118\x}
                \tikzset{xshift = \xx * 0.6cm}
                \foreach \y in {1, ..., 5} {
                    \pgfmathrandomitem\A{ACGT}
                    \tikzset{yshift = \y * 0.5cm}
                    \draw [\A] (0, 0.4) rectangle (0.1, 0)
                        node [base] {\A} (0.5, 0.4);
                }
            }
        }
    \end{tikzpicture}
    \caption{
        The problem Geno-Weaving~\cite{WaG25} tries to solve:
        Some strands are sampled once and
        some are sampled multiple times.
        If we apply error-correcting codes vertically,
        single-read strands provide too little information
        while multi-read strands provide too much.
        By coding horizontally, the read multiplicity
        becomes part of the channel randomness.
    }                                                       \label{fig:weave}
\end{figure}

    Naturally, one might design a DNA coding scheme by first
    designing codes specifically for insertion and deletion on a strand
    and then applying an optional outer code across all strands,
    as illustrated in Figure~\ref{fig:concat}.
    However, DNA storage has a distinctive structure: data is encoded
    across a large ($2^8$--$2^{20}$) pool of short ($2^6$--$2^8$)
    strands \cite{RGC26}, suggesting an alternative approach.
    In our recent work~\cite{WaG25}, we proposed Geno-Weaving,
    a framework that achieves the channel capacity with practical
    time complexity by ``coding in the orthogonal direction'':
    protecting bits across
    strands rather than along individual strands,
    as illustrated in Figure~\ref{fig:weave}.
    This approach converts randomness in the number of reads per
    strand into substitution-like noise, enabling the use of
    well-established codes for substitution channels.
    Building on Geno-Weaving, we recently extended this framework to
    handle deletion errors~\cite{LWG25} by incrementally aligning
    the received sequences and correcting errors symbol-by-symbol.
    However, that work was limited to deletion-only channels.

    In this paper, we resolve this limitation by introducing a
    \emph{salami slicing trellis} that accounts for substitution,
    insertion, and deletion patterns when computing bitwise posterior
    probabilities, as shown in Figure~\ref{fig:salami}.
    Inspired by trellis decoders in related works \cite{MLW22,TPF22,ArT23},
    our approach maintains the relevant state space,
    allowing the decoder to make optimal estimates
    in the presence of all error types.
    The result is a unified framework that can handle
    $1\%$ errors of each type while approaching
    the conjectured capacity.

    The paper is organized as follows.
    Section~\ref{sec:problem} formalizes the problem setting.
    Section~\ref{sec:geno} reviews Geno-Weaving,
    a technique in which coding in the orthogonal direction
    converts channel randomness into substitution-like noise.
    Section~\ref{sec:salami} introduces the salami slicing trellis.
    Section~\ref{sec:ortho} explains how to
    integrate the salami slicing trellis with orthogonal codes.

\section{Problem Statement}                               \label{sec:problem}

    In this section we state the mathematical model.
    A DNA pool has $n$ strands, where $n \in [2^8, 2^{20}]$.
    These strands have length $\ell$, where $\ell \in [2^6, 2^8]$,
    and are denoted by $X^{(1)}, \dotsc, X^{(n)} \in \{0, 1\}^\ell$.
    The individual bits of $X^{(s)}$ are denoted by
    $X^{(s)}_1, \dotsc, X^{(s)}_\ell \in \{0, 1\}$.

    The sequencer reads each strand exactly once,
    and outputs $Y^{(1)}, \dotsc, Y^{(n)}$.
    Entries of $Y^{(s)}$ are denoted by
    $Y^{(s)}_1$, $\dotsc, Y^{(s)}_{|Y^{(s)}|} \in \{0, 1\}$,
    where $|Y^{(s)}|$ is a random variable that depends on
    $\ell$ and on the number of insertions and deletions.
    The complement of $Y^{(s)}_q$ is denoted by
    $\bar Y^{(s)}_q \coloneqq 1 - Y^{(s)}_q$.
    
    The generation of $Y^{(s)}$ from $X^{(s)}$
    is modeled by a triply-noisy channel with
    \begin{itemize}
        \item substitution probability $\varsigma \in [0, 1\%]$,
        \item insertion probability $\iota \in [0, 1\%]$, and
        \item deletion probability $\delta \in  [0, 1\%]$.
    \end{itemize}
    Their complements are denoted by
    $\bar\varsigma$, $\bar\iota$, and $\bar\delta$, respectively.
    To be more precise,
    each bit $X^{(s)}_p$ first determines if it is deleted;
    if not, it determines if it is substituted (flipped).
    For each gap\footnote{
    The gap before the first bit and the gap after the last bit
    are also gaps.}
    between bits, we consider memoryless insertions.
    Thus the number of inserted bits follows a geometric distribution
    whose mean is $\iota / \bar\iota$,
    and the inserted bits are iid uniform.
    All deletion, substitution, and gap-insertion trials are independent
    across strands and positions.

    The problem is to design a coding scheme to encode as much information
    as possible into $X^{(1)}, \dotsc, X^{(n)}$ that can be decoded from
    $Y^{(1)}, \dotsc, Y^{(n)}$ with a small enough probability of error.
    Note that the present paper considers the setting in which
    each strand produces exactly one read and the reads are not shuffled.
    This simplification lets us isolate and attack
    the desynchronization challenge.
    For works that consider both synchronization
    and multiple reads of a single strand, see \cite{CGM19,BLS20};
    for works that consider synchronization and multiple reads
    of multiple strands, see \cite{MLW22,SGP24}.

    In the next section, we recall Geno-Weaving in its original
    randomly-many-reads setting, because the same idea of coding
    across strands will be reused for this single-read model.

\section{Geno-Weaving}                                       \label{sec:geno}

    In previous work~\cite{WaG25}, we proposed Geno-Weaving,
    the first low-complexity coding scheme that achieves
    the DNA coding capacity determined by a long series of
    works \cite{HSR17,LSW19,LSW20,ShH21,LHS22,WeM22,LSW23}.
    The channel model studied by these works is the combination
    of sampling randomness and substitution errors:
    Each strand $X^{(s)}$ will be sampled $K^{(s)}$ times,
    where $K^{(s)}$ is a Poisson random variable,
    and each $Y^{(s,k)}$, $k \in [K^{(s)}]$,
    is generated from $X^{(s)}$ by a substitution channel, say $\BEC(1/2)$.

    To design a low-complexity coding scheme for this channel,
    a natural approach is to apply error-correcting codes along each strand,
    and protect the pool of strands by an outer code, as illustrated in
    Figure~\ref{fig:concat} (cf.\ \cite{WML23,MRA23,MLW22}).
    That is, find an inner code $\III \subset \{0, 1\}^\ell$
    and make $X^{(s)} \in \III$ for all $s \in [n]$.
    Then find an outer code $\OOO \subset \III^n$
    and make $(X^{(1)}, \dotsc, X^{(n)}) \in \OOO$.
    This, however, does not achieve the channel capacity.
    To see why, notice that reading $X^{(s)}$ $K^{(s)}$ times via
    $\BEC(1/2)$ is equivalent to reading $X^{(s)}$ once
    via $\BEC(2^{-K^{(s)}})$.
    In other words, the channel quality varies from strand to strand,
    making it impossible to select one inner code rate
    to match the channel capacities of all strands.
    
    The key idea developed by Geno-Weaving~\cite{WaG25}
    is shown in Figure~\ref{fig:weave}:
    protect the $p$th position of the strands using a block code,
    meaning that $(X^{(1)}_p, \dotsc, X^{(n)}_p) \in \CCC$ for some
    block code $\CCC \subset \{0, 1\}^n$ and for all $p \in [\ell]$.
    Now each $X^{(s)}_p$ will be read $K^{(s)}$ times,
    so $Y^{(s)}_p$ can be seen as the output of $\BEC(p)$, where
    $p = \sum_{k=0}^\infty \frac{1}{2^k} \PP(K^{(s)}{=}k) = e^{-\lambda/2}$
    and $\lambda$ is the mean of $K^{(s)}$.
    Since $K^{(s)}$ varies from symbol to symbol, these BECs are independent,
    which implies that we can let $\CCC$ be any
    capacity-achieving code for $\BEC(e^{-\lambda/2})$.
    To summarize, Geno-Weaving codes in the orthogonal direction
    to convert the randomness in the number of reads
    into substitution-like noise, and the latter
    can be tackled by well-studied codes for substitution channels.

\begin{figure}
    \centering
    \begin{tikzpicture}
        \footnotesize
        \foreach \y in {1, ..., 5} {
            \tikzset{yshift = \y * 0.5cm}
            \draw [rounded corners=3pt] (0.8, -0.05) rectangle (7.7, 0.4)
                node [below right] {polar code};
        }
        \foreach \x in {1, ..., 7} {
            \tikzset{xshift = \x * 1cm}
            \draw [rounded corners=3pt] (-0.1, 3.1) rectangle (0.5, 0.3)
                (0.2, 0.3) node [below] {Fig.~\ref{fig:trellis}};
        }
        \normalsize
        \foreach \x in {1, ..., 7} {
            \tikzset{xshift = \x * 1cm}
            \foreach \y in {1, ..., 5} {
                \pgfmathrandomitem\A{ACGT}
                \tikzset{yshift = \y * 0.5cm}
                \draw [\A] (0, 0.4) rectangle (0.1, 0)
                    node [base] {\A} (0.5, 0.4);
            }
        }
    \end{tikzpicture}
    \caption{
        The overall design of our scheme.
        Each strand maintains a salami slicing trellis
        (Figure~\ref{fig:trellis}).
        Each position is protected by a polar code.
    }                                                      \label{fig:salami}
\end{figure}
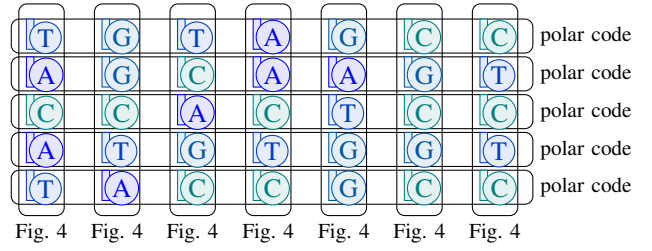

    In a more recent work~\cite{LWG25}, we discuss an extension of
    Geno-Weaving to the case where strands may encounter deletion errors.
    Without Geno-Weaving, the naive approach to handle
    deletion errors is again to apply a block code along each strand,
    $X^{(s)} \in \III$ for all $s \in [n]$,
    and protect the pool of strands by an outer code,
    $(X^{(1)}, \dotsc, X^{(n)}) \in \OOO$.
    This approach does achieve the capacity when $\ell \to \infty$
    because there are codes that achieve
    deletion capacity \cite{TPF22,ArT23,PLW22}.
    However, for DNA coding the length of each strand
    is limited by synthesis technologies \cite{TJL24},
    and the finite-length effect~\cite{PPV10} penalizes the code rate.
    But by coding in the orthogonal direction,
    i.e., $(X^{(1)}_p, \dotsc, X^{(n)}_p) \in \CCC$ for all $p \in [\ell]$,
    the block length of $\CCC$ is $n$, which is about the square of $\ell$.
    We term this the \emph{block length gain}.
    Coding in the orthogonal direction also means that
    any deletion error will look like a substitution error to the decoder,
    and we are able to take advantage of the rich literature on
    substitution-correcting codes.

    A limitation of~\cite{LWG25} is that it can handle deletion errors only.
    This is because whenever the decoder detects a deletion,
    it inserts the deleted bit back before proceeding to the next $p$.
    If both insertion and deletion errors exist,
    the decoder cannot determine whether it should insert
    the missing bit back or delete the extra bit.
    To resolve this, we use a trellis that computes
    the precise posterior distribution of the next bit
    by taking both insertion and deletion probabilities into account.

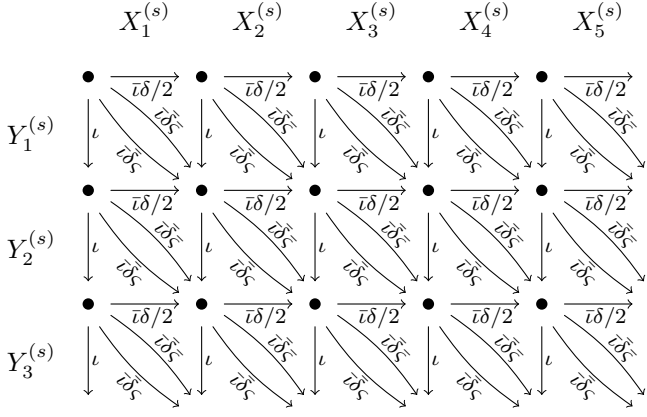
\begin{figure}
    \centering
    \begin{tikzpicture} [scale=1.5, inner sep=1pt]
        \def\ll{5}
        \def\mm{3}
        \foreach \x in {1, ..., \ll} {
            \draw (\x.5, -0.5) node {$X^{(s)}_\x$};
        }
        \foreach \y in {1, ..., \mm} {
            \draw (0.5, -\y.5) node {$Y^{(s)}_\y$};
        }
        \footnotesize
        \foreach \x in {1, ..., \ll} {
            \foreach \y in {1, ..., \mm} {
                \fill (\x, -\y) circle (0.5mm);
                \draw [->] (\x, -\y) +(0, -0.2) --
                    node [right] {$\iota$} +(0, -0.8);
                \draw [->] (\x, -\y) +(0.2, 0) --
                    node [below] {$~~\bar\iota \delta/2$} +(0.8, 0);
                \draw [->] (\x, -\y) +(0.1, -0.2) to [bend right=10]
                    node [below, sloped]
                    {~~$\bar\iota \bar\delta \varsigma$} +(0.8, -0.9);
                \draw [->] (\x, -\y) +(0.2, -0.1) to [bend left=10]
                    node [above, sloped]
                    {~~$\bar\iota \bar\delta \bar\varsigma$} +(0.9, -0.8);
            }
        }
    \end{tikzpicture}
    \caption{
        The salami slicing trellis.
        Columns and rows are numbered from $0$.
        To estimate the posterior of $X^{(s)}_1$,
        column $1$ is computed.
        Then $X^{(s)}_1$ is revealed by the oracle,
        and column $2$ is computed to estimate $X^{(s)}_2$.
        Then $X^{(s)}_2$ is revealed
        for the computation of column $3$, and so on.
    }                                                     \label{fig:trellis}
\end{figure}

\section{Salami Slicing Trellis}                           \label{sec:salami}

    The idea of using a trellis is taken from the related work
    \cite[Figure~3]{MLW22} as well as polar coding for deletion
    channels~\cite[Figure~2]{TPF22}~\cite[Figure~5]{ArT23}.
    There, they build a trellis that can keep track of
    the posterior probabilities of different error patterns.
    They then apply polar transformations to the trellis
    to split it into two smaller trellises,
    and repeat recursively until the trellis is small enough
    that brute-force decoding is possible.

    In this work, we build an interactive trellis
    that we call a \emph{salami slicing trellis}.
    Before we can describe how the trellis is integrated
    in the overall decoder, we first explain how one trellis works
    on a single strand in the following subsection.

\subsection{An Interactive Trellis}

    Fix a strand index $s \in [n]$.
    The trellis associated with this strand sees the entire output $Y^{(s)}$
    and is tasked to compute the posterior distribution
    of the first bit $X^{(s)}_1$.
    After this is done for all $s \in [n]$,
    an oracle tells the trellis the true value of $X^{(s)}_1$.
    It is then tasked to compute the posterior distribution
    of the second bit $X^{(s)}_2$.
    Afterward the true value of $X^{(s)}_2$ is revealed by the oracle.
    And the trellis computes the posterior distribution of $X^{(s)}_3$.
    The procedure continues alternating between
    estimating $X^{(s)}_p$ and learning the true value of $X^{(s)}_p$
    from the oracle until $p$ reaches $\ell$.

    The overall structure of the trellis
    is depicted in Figure~\ref{fig:trellis}.
    The trellis has $\ell + 1$ columns numbered $0, \dotsc, \ell$.
    For $p \in [\ell]$, the gap between columns $p - 1$ and $p$
    corresponds to the true bit $X^{(s)}_p$.
    It has $|Y^{(s)}| + 1$ rows numbered $0, \dotsc, |Y^{(s)}|$.
    For $q \in [|Y^{(s)}|]$, the gap between rows $q - 1$ and $q$
    corresponds to the observation $Y^{(s)}_q$.
    A rightward arrow\footnote{
    Strictly speaking, there are two deletion arrows:
    one that deletes $0$ and one that deletes $1$;
    each has probability $\bar\iota\delta/2$.
    When computing the posterior probabilities of,
    say, $X^{(s)}_p = Y^{(s)}_q$,
    exactly one of these two arrows is applicable.
    Therefore we draw a single rightward arrow labeled $\bar\iota\delta/2$.}
     corresponds to a deletion event,
    where a bit $X^{(s)}_p$ is deleted and does not go into $Y^{(s)}$.
    A downward arrow corresponds to an insertion event,
    where an extra bit $Y^{(s)}_q$ is inserted.
    The diagonal arrows correspond to substitution or no-error events.
    The $\varsigma$-arrow applies if $X^{(s)}_p$ is not equal to $Y^{(s)}_q$,
    and the $\bar\varsigma$-arrow applies if they are equal.

    Next we start estimating $X^{(s)}_1$.
    The probability of $X^{(s)}_1 = 1$ can be decomposed into three cases:
    \begin{itemize}
        \item $X^{(s)}_1$ is deleted.
            Half of the time it is $1$ and half of the time it is $0$;
            this contributes $\bar\iota\delta/2$.
        \item $X^{(s)}_1$ is substituted; this contributes
            $\bar\iota\bar\delta \varsigma \bar Y^{(s)}_1$.
        \item $X^{(s)}_1$ is kept; this contributes
            $\bar\iota\bar\delta \bar\varsigma Y^{(s)}_1$.
    \end{itemize}
    Their sum is
    $\bar\iota\delta/2
    + \bar\iota\bar\delta \varsigma \bar Y^{(s)}_1
    + \bar\iota\bar\delta \bar\varsigma Y^{(s)}_1$.
    It can also be the case that the
    first observation $Y^{(s)}_1$ is an insertion,
    and hence we should use the second observation $Y^{(s)}_2$
    to help estimate $X^{(s)}_1$.
    Sometimes, even $Y^{(s)}_2$ is an insertion
    and we should look at $Y^{(s)}_3$.
    Continuing this logic, we have to sum over all $q$ to get
    \[
        \PP \bigl\{ X^{(s)}_1{=}1 \mid Y^{(s)} \bigr\} = \frac1Z
        \sum_q \iota^{q-1} \Bigl(
            \frac{\bar\iota\delta}2
            + \bar\iota\bar\delta \varsigma \bar Y^{(s)}_q
            + \bar\iota\bar\delta \bar\varsigma Y^{(s)}_q
        \Bigr),
    \]
    where $Z \coloneqq \sum_{q=1}^{|Y^{(s)}|} \iota^{q-1}\bar\iota$
    is the normalization factor.

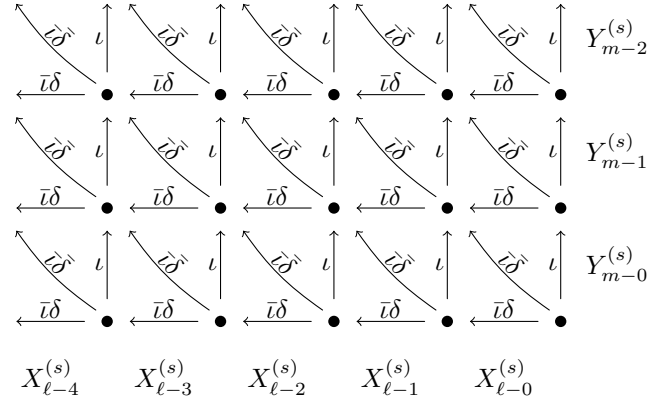
\begin{figure}
    \centering
    \begin{tikzpicture} [scale=1.5, inner sep=1pt]
        \def\ll{4}
        \def\mm{2}
        \foreach \x in {0, ..., \ll} {
            \draw (-\x.5, -0.5) node {$X^{(s)}_{\ell-\x}$};
        }
        \foreach \y in {0, ..., \mm} {
            \draw (0.5, \y.5) node {$Y^{(s)}_{m-\y}$};
        }
        \foreach \x in {0, ..., \ll} {
            \foreach \y in {0, ..., \mm} {
                \fill (-\x, \y) circle (0.5mm);
                \draw [->] (-\x, \y) +(0, 0.2) --
                    node [left] {$\iota$} +(0, 0.8);
                \draw [->] (-\x, \y) +(-0.2, 0) --
                    node [above] {$\bar\iota \delta$} +(-0.8, 0);
                \draw [->] (-\x, \y) +(-0.1, 0.1) to [bend left=10]
                    node [above, sloped] {$\bar\iota \bar\delta$~~}
                    +(-0.8, 0.8);
            }
        }
    \end{tikzpicture}
    \caption{
        The tail trellis provides a correction term
        for the posterior distribution obtained from
        the main trellis in Figure~\ref{fig:trellis}.
        The goal is to compute the likelihood that the tail of $Y^{(s)}$
        is generated by the tail of $X^{(s)}$ and insertions/deletions.
        Hence no $\varsigma$ is involved
        and the $\delta$-arrows do not have the $1/2$ factor.
    }                                                        \label{fig:tail}
\end{figure}

    To carry out this computation efficiently,
    use the trellis in Figure~\ref{fig:trellis}:
    Start from putting a $1$ on the top left node,
    which is column $0$ and row $0$,
    follow the arrows and multiply the probabilities along the way.
    Note that one has to choose either the $\varsigma$-arrow
    or the $\bar\varsigma$-arrow depending on the corresponding $Y$ value.
    Then the posterior probability of $X^{(s)}_1 = 1$
    is the sum of the values on column $1$.

    We now proceed to the second bit $X^{(s)}_2$.
    Let $r$ be the index such that
    $Y^{(s)}_1, \dotsc, Y^{(s)}_r$
    are generated by $X^{(s)}_1$ or by insertions before.
    Then the posterior probability of $X^{(s)}_2 = 1$
    can be decomposed into three cases:
    \begin{itemize}
        \item $X^{(s)}_2$ is deleted.
            Half of the time it is $1$ and half of the time it is $0$;
            this will contribute $\bar\iota\delta/2$.
        \item $X^{(s)}_2$ is substituted; this contributes
            $\bar\iota\bar\delta \varsigma \bar Y^{(s)}_{r+1}$.
        \item $X^{(s)}_2$ is kept; this contributes
            $\bar\iota\bar\delta \bar\varsigma Y^{(s)}_{r+1}$.
    \end{itemize}
    Again, we have to take into account
    that $Y^{(s)}_{r+1}$ may be an insertion,
    and hence we should look at $Y^{(s)}_{r+2}$, $Y^{(s)}_{r+3}$,
    and so on.
    Therefore, the posterior probability of $X^{(s)}_2 = 1$ is
    \[
        \frac1Z
        \sum_r L_r \sum_q \iota^{q-1} \Bigl(
            \frac{\bar\iota\delta}2
            + \bar\iota\bar\delta \varsigma \bar Y^{(s)}_{r+q}
            + \bar\iota\bar\delta \bar\varsigma Y^{(s)}_{r+q}
        \Bigr),
    \]
    where $L_r$ is the likelihood that $Y^{(s)}_1, \dotsc, Y^{(s)}_r$
    are generated by $X^{(s)}_1$ or by insertions before, and $Z$
    is the normalization factor
    $\sum_r L_r \sum_q \iota^{q-1}\bar\iota$.

    To carry out this computation efficiently, notice that $L_r$
    is the value associated with the node in row $r$ of column $1$,
    with one major difference:
    We now know the correct value of $X^{(s)}_1$, so we compute\footnote{
        $|X - Y|$ is the indicator of $X \ne Y$, and
        $\overline{|X - Y|}$ is the indicator of $X = Y$.
    }
    $\bar\iota\bar\delta \varsigma |X^{(s)}_1 - Y^{(s)}_q|
    + \bar\iota\bar\delta \bar\varsigma
    \overline{|X^{(s)}_1 - Y^{(s)}_q|}$
    instead of 
    $\bar\iota\bar\delta \varsigma \bar Y^{(s)}_q
    + \bar\iota\bar\delta \bar\varsigma Y^{(s)}_q$.
    That is, we have to choose between the $\varsigma$-arrow
    and the $\bar\varsigma$-arrow
    according to whether the true value of $X^{(s)}_1$
    matches the corresponding $Y$ value.
    With this, the posterior probability of $X^{(s)}_2 = 1$
    should be the sum of column $2$ up to the normalization factor.

    The same process can be repeated for all $p = 1, \dotsc, \ell$:
    We follow the arrows to fill in the values of the nodes
    on column $p$, and then sum over the entire column
    to get the posterior probability of $X^{(s)}_p = 1$.
    After a column is processed,
    the oracle reveals the true value of $X^{(s)}_p$,
    which will be put back to the trellis
    for the computation of the $L$'s for the next column.

\subsection{The Tail Trellis}

    The main trellis depicted in Figure~\ref{fig:trellis}
    and explained in the previous subsection
    can compute the posterior probabilities very well
    when $p$ and $q$ are relatively small compared to $\ell$.
    In particular, the value at column $p$ and row $q$
    is supposed to be the probability
    that $Y^{(s)}_1, \dotsc, Y^{(s)}_q$ are generated by
    $X^{(s)}_1, \dotsc, X^{(s)}_p$ and insertions/deletions before.
    This, however, does not consider the fact that
    $X^{(s)}_{p+1}, \dotsc, X^{(s)}_\ell$
    and insertions/deletions after $X^{(s)}_{p+1}$ need to generate
    $Y^{(s)}_{q+1}, \dotsc, Y^{(s)}_{|Y^{(s)}|}$ as well.
    That is to say, if $(1 + \iota - \delta) (\ell - p)$
    is not approximately equal to $|Y^{(s)}| - q$,
    then such a deviation has small probability.

    To properly penalize the $(p, q)$-pairs whose tail part is rare,
    we duplicate and modify the main trellis by merging each
    $\varsigma$-arrow and $\bar\varsigma$-arrow into a single diagonal arrow.
    We also change $\delta / 2$ to $\delta$ for the rightward arrows
    because the tail part does not care about the bit values.
    The result is depicted in Figure~\ref{fig:tail}.
    This trellis will then compute the probability that the tail of $Y^{(s)}$
    is generated by the tail of $X^{(s)}$ and insertions/deletions.
    We multiply the values from the main trellis and the tail trellis
    to get the posterior probability of $X^{(s)}_p = 1$.

    Complexity-wise, the time spent on the main trellis
    is linear in the number of arrows, which is $O(\ell^2)$.
    The same applies to the tail trellis.
    We infer that the overall complexity of the salami slicing trellis
    is $O(\ell^2)$ per strand, or $O(n \ell^2)$ per pool.
    In the complete decoder, this is combined with one polar decoding
    operation of length $n$ for each of the $\ell$ positions.
    With successive cancellation decoding, the polar part therefore costs
    $O(\ell n \log n)$, so the total decoding complexity is
    $O(n \ell^2 + \ell n \log n)$ per pool.
    By the same deviation argument before,
    one sees that paths too far away from the diagonal
    (whose slope is about $1 + \iota - \delta$)
    will have exponentially small probabilities,
    and hence contribute little to the posterior probabilities.
    This could help reduce the complexity.

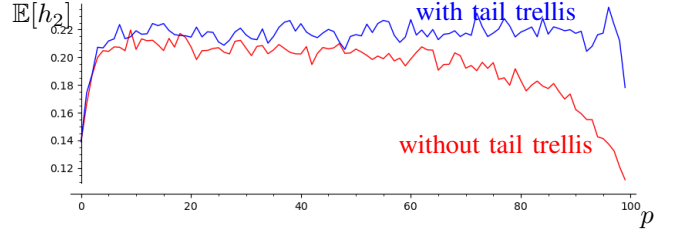
\begin{figure}
    \centering
    \includegraphics[width=8cm]{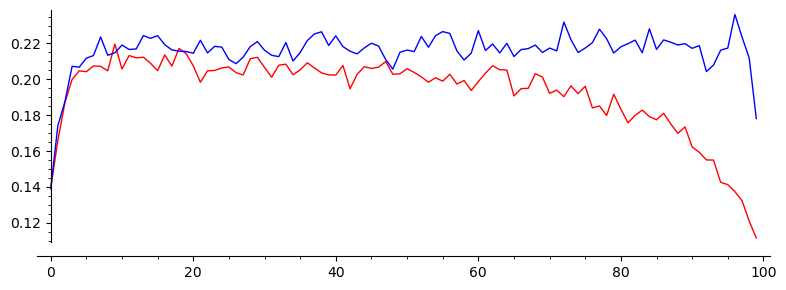}%
    \tikz [overlay] {
        \draw(-8, 3) node [below] {$\EE[h_2]$};
        \draw(0, 0) node {$p$};
        \draw(-2, 3) node [below, blue] {with tail trellis};
        \draw(-2, 1) node [red] {without tail trellis};
    }
    \caption{
        Empirical average of $1000$ $h_2$ values (vertical axis)
        at different position $p$ (horizontal axis);
        $\ell = 100$ and $\varsigma =\iota = \delta = 1\%$.
        The upper blue curve does not consider the tail trellis,
        while the lower red curve does.
        The average $h_2$ values are $0.216$ for the former
        and $0.194$ for the latter.
        The conjectured equivocation is $3 h_2(1\%) = 0.242$.
    }                                                          \label{fig:h2}
\end{figure}

\section{Orthogonal Polar Decoding}                       \label{sec:ortho}

    The salami slicing trellis turns each position of each strand into
    a posterior probability.
    This section explains how these posterior probabilities are used.
    We first study the empirical channel seen by an ideal oracle, and
    then replace that oracle by a polar code applied in the orthogonal
    direction.

\subsection{The Channel the Oracle Sees}

    The posterior probabilities, one for each
    $(p, s) \in [\ell] \times [n]$, will be used by polar codes.
    But for now, let us still assume that
    the posterior probabilities are fed into an oracle.
    The oracle needs to reveal the corrected value of $X^{(s)}_p$
    to the trellis upon receiving the posterior probability,
    so we are interested in the entropy it spends
    to turn an estimate to the true value.
    In other words, we are interested in studying
    \[
        T^{(s)}_p\colon X^{(s)}_p
        \to \text{the $p$th output of the $s$th trellis}
    \]
    the trellis-output channel.

    It is difficult to describe $T^{(s)}_p$ analytically.
    But we can fall back to the sampling approach:
    First, notice that $T^{(s)}_p$ does not depend on $s$
    because the strands are all treated the same.
    We then sample, say, $1000$ strands and run the salami slicing trellis
    on them to get $1000$ posterior probabilities for each $p$.
    With this we know that
    \[
        1 - \frac1{\text{\#samples}} \sum h_2(\text{posterior probability})
    \]
    is an empirical estimate of the capacity of $T^{(s)}_p$,
    where $h_2$ is the binary entropy function.
    See Figure~\ref{fig:h2} for a plot of this estimate for each $p$.
    
    In our simulations, the empirical capacity is close to
    \[ 1 - h_2(\varsigma) - h_2(\iota) - h_2(\delta). \]
    The heuristic for the formula is as follows:
    For BSC with small $\varsigma$, the output string
    will share long segments with the input string,
    and symbols that do not belong to these segments
    are the ones flipped by the channel.
    The receiver learns the locations of these flips,
    and so the capacity penalty of BSC is $h_2(\varsigma)$.
    The same heuristic holds for
    binary deletion channel (BDC) with small $\delta$:
    We can match long segments of the input and output
    strings to learn the locations of deletions,
    and hence the penalty is still $h_2(\delta)$~\cite{KMS10}.
    Their combination is discussed in~\cite{KaD25},
    where the authors discuss small substitution and small deletion errors,
    and the penalty is $h_2(\varsigma) + h_2(\delta)$.

    From what is avbve ww conjecture that for a triply-noisy channel
    with small $\varsigma$, $\iota$, and $\delta$,
    the capacity is $1 - h_2(\varsigma) - h_2(\iota) - h_2(\delta)$.
    This matches the number we get in Figure~\ref{fig:h2}.
    For more discussion on the capacity, see \cite{FDE11,RaD13,MLR23,MRA23}.

\subsection{Polar Code as the Oracle}

    We now explain how to construct and use a polar code~\cite{Ari09}
    to serve as the oracle that reveals the true value of $X^{(s)}_p$
    to the trellis.
    Note that the polar code can make mistakes, so it is not a perfect oracle.
    But we still pretend that,
    when the trellis is computing the posterior distribution
    at location $p + 1$, the values of $X^{(s)}_1, \dotsc, X^{(s)}_p$
    revealed to the trellis by the polar code are correct.
    This is because, if they are not correct,
    then the current pool already has a decoding failure,
    and anything we do later will not recover from it anyway.
    This ``assume previous decisions are correct'' principle
    is the same one used in the successive cancellation decoder
    for polar codes~\cite{Ari09}.

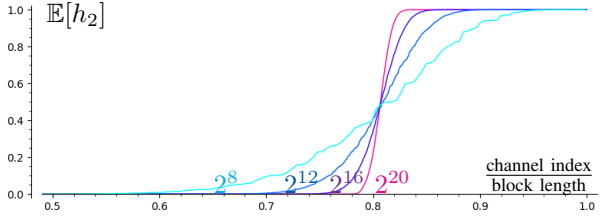
\begin{figure}
    \includegraphics[width=8cm]{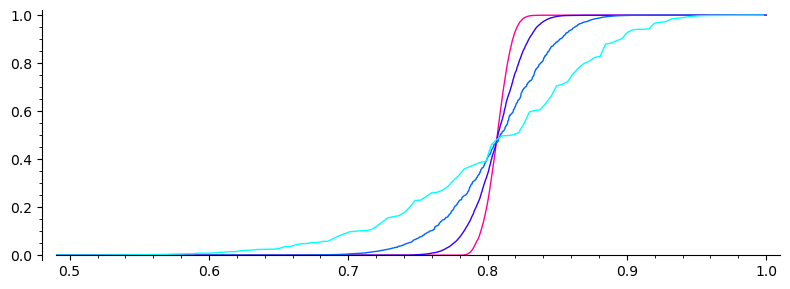}%
    \tikz [overlay] {
        \draw(-7.5, 3) node [below right] {$\EE[h_2]$};
        \draw(0, 0.2) node [above left]
            {$\frac{\text{channel index}}{\text{block length}}$};
        \draw(-5, 0.5) node [cyan] {$2^8$};
        \draw(-4, 0.5) node [cyan!30!blue] {$2^{12}$};
        \draw(-3.4, 0.5) node [magenta!30!blue] {$2^{16}$};
        \draw(-2.8, 0.5) node [magenta] {$2^{20}$};
    }
    \caption{
        The empirical capacities of the bit channels after sorting;
        $\ell = 100$ and $\varsigma =\iota = \delta = 1\%$.
        From cyan (gentle) to magenta (steep), the curves correspond to
        $n = 2^8, 2^{12}, 2^{16}, 2^{20}$, respectively.
    }                                                   \label{fig:waterfall}
\end{figure}

    As mentioned above, we can simulate the trellis for as many strands
    as we want and collect the posterior probabilities.
    This defines an empirical approximation of the true channel $T^{(s)}_p$
    that can be used to construct polar codes.
    We apply the standard polar transformation to the empirical channels
    to get the empirical version of the bit channels,
    and estimate their capacities thereby.
    We then sort the bit channels according to their capacities
    to see if (a) the channels polarize as the block length grows,
    and (b) the fraction of good channels approaches
    the (conjectured) capacity of $T^{(s)}_p$.

    The results are plotted in Figure~\ref{fig:waterfall}.
    There, we let $\varsigma, \iota, \delta$ all be $1\%$.
    We can see that the thresholds are very close to the conjectured value
    $1 - h_2(\varsigma) - h_2(\iota) - h_2(\delta) = 0.758$.

\subsection{Simulation of Whole Pools}

    We now explain the overall design of our coding scheme that combines
    the salami slicing trellis and polar codes.
    The overall design is depicted in Figure~\ref{fig:salami}.
    The encoder is as follows.
    \begin{itemize}
        \item For each $p \in [\ell]$, construct a polar code
            targeting $T^{(s)}_p$.
        \item Encode messages to produce
        $X^{(s)}_p$, $s \in [n]$, $p \in [\ell]$.
        \item Synthesize the DNA pool with strands $X^{(s)}$, $s \in [n]$.   
    \end{itemize}
    The decoder alternates between the salami slicing trellis
    and the polar code $\ell$ times;
    start with $p = 1$:
    \begin{itemize}
        \item After the $p$th iteration, compute the posterior distributions
            of $X^{(s)}_p$, $s \in [n]$ using the trellis.
        \item Decode the polar code for the $p$th position
            to obtain the corrected values of $X^{(s)}_p$, $s \in [n]$.
            Here we use the successive cancellation decoder.
        \item Feed the decoded $X^{(s)}_p$
            back to the salami slicing trellis.
    \end{itemize}
    
\begin{figure}
    \includegraphics[width=8cm]{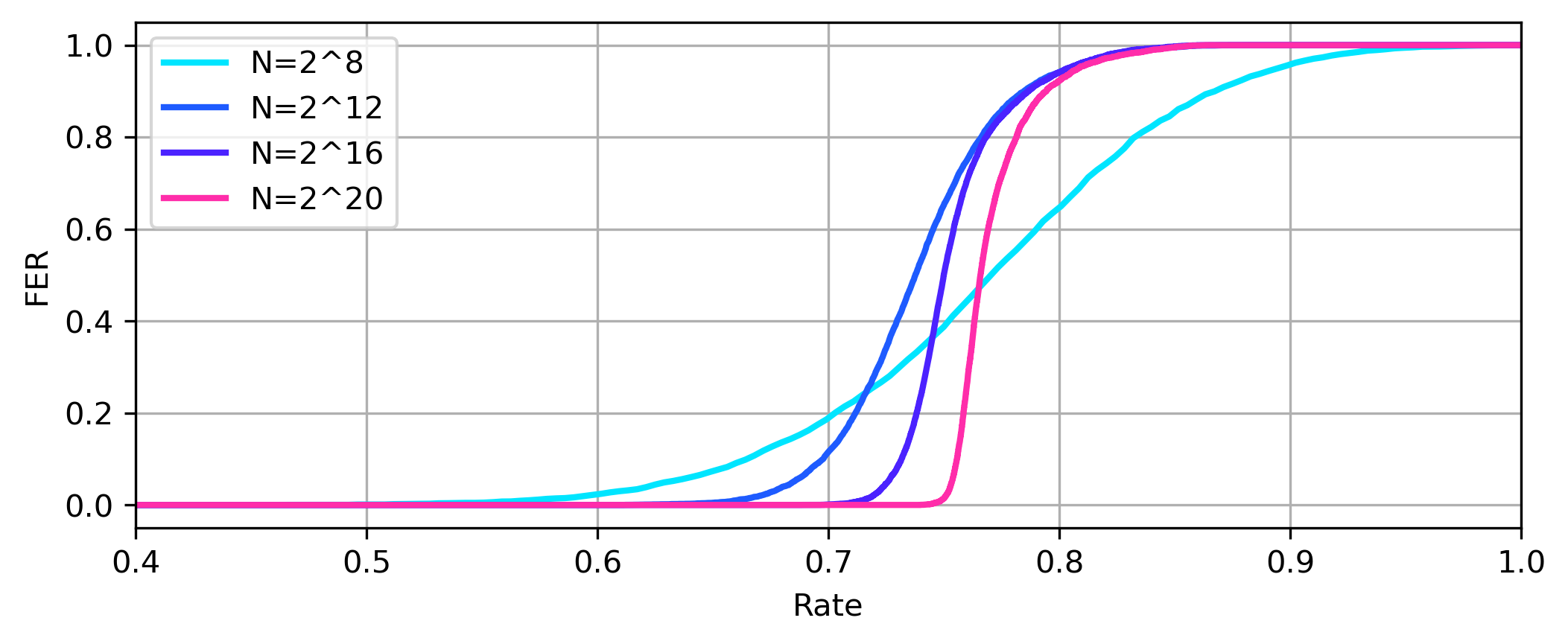}%
    \tikz [overlay] {
    }
    \caption{
        The block error probability.
        ($100$ blocks per pools; $100$ pools.)
    }                                                         \label{fig:fer}
\end{figure}

\begin{figure}
    \includegraphics[width=8cm]{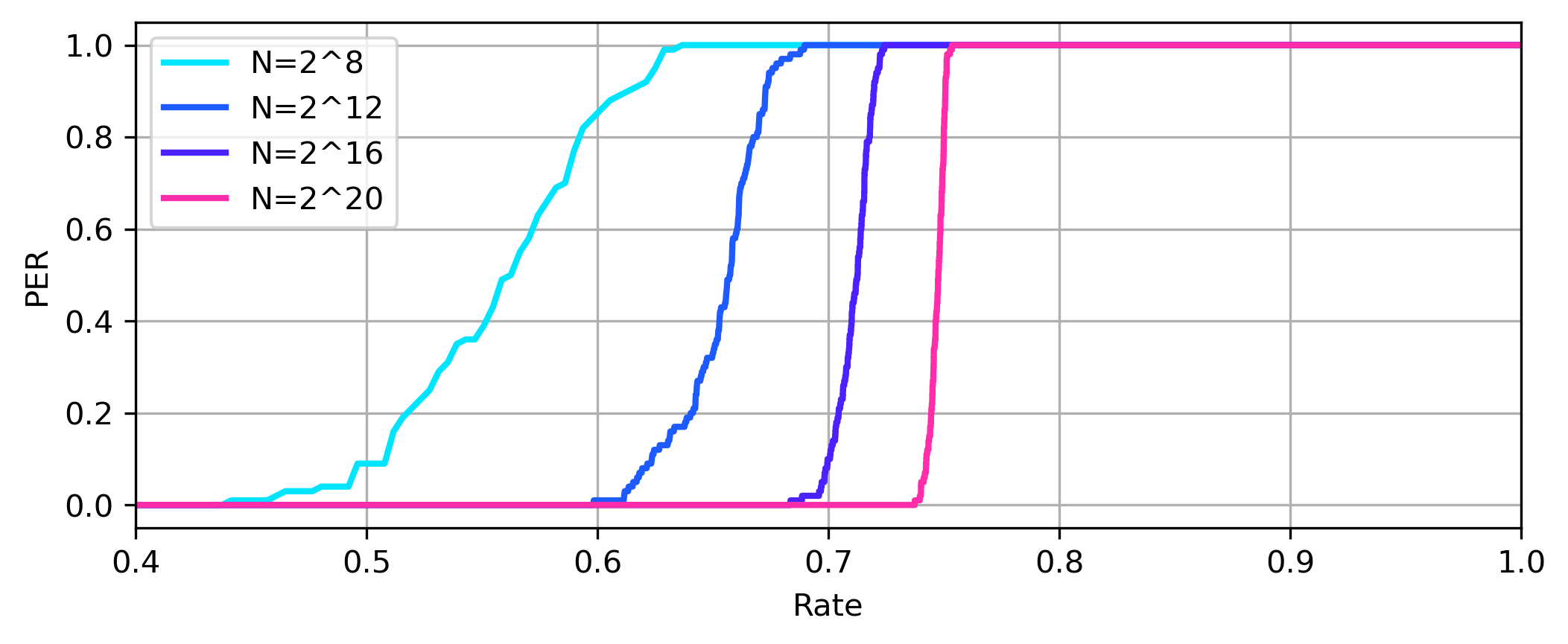}%
    \tikz [overlay] {
    }
    \caption{
        The result of repeating the expriment for
        $(\varsigma, \iota, \delta) = (1\%, 2\%, 3\%)$
    }                                                         \label{fig:per}
\end{figure}

    We then simulate $20$ pools, each containing $n$ strands and $\ell =
    20$ bits per strand, where $n \in \{2^8, 2^{12}, 2^{16}, 2^{20}\}$.
    This creates $n \ell$ bit channels per pool, each with $20$ samples.
    We select the bit chnnels with the highest $h_2$ average.
    We then simulate $100$ pools for each $n$.
    Against code rate, we plot the block error probability
    (one block is $n$ bits) in Figure~\ref{fig:fer}
    and the pool error probability
    (one pool is $\ell$ blocks) in Figure~\ref{fig:per}.
    Observe that the code rates does approach the (conjectured) capacity
    as $n$ grows.

    % Additionally, we repeat the same experiment six more times with
    % $\{\varsigma, \iota, \delta\} = \{1\%, 2\%, 3\%\}$.
    % The results are gathered in Figures ...
    % The results also support the earlier claim that as $n \to \infty$,
    % the overall code rate increases by a considerable amount.
    % They also support the conjecture that
    % the channel capacity is close to, if not equal to,
    % $1 - h_2(\varsigma) - h_2(\iota) - h_2(\delta) \approx 0.583$.

\section{Conclusion}

    In this paper, we introduce the salami slicing trellis,
    which can compute the posterior distributions of bits
    transmitted over a triply-noisy channel with
    substitution, insertion, and deletion errors.
    By combining the salami slicing trellis with polar codes
    applied in the orthogonal direction,
    we construct a coding scheme that approaches
    the conjectured capacity of
    $1 - h_2(\varsigma) - h_2(\iota) - h_2(\delta)$
    for small error probabilities $\varsigma$, $\iota$, and $\delta$.

\bibliographystyle{IEEEtran}
\bibliography{SalamiSlicing-30}

\end{document}